\documentclass[table]{ccjnl}
\usepackage{lipsum,amsmath}
\usepackage{cuted}
\usepackage{bm}
\usepackage{verbatim}
\usepackage{makecell}
\usepackage{footnote}
\usepackage{color}

\graphicspath{{figures/}}

\title{The First Verification Test of Space-Ground Collaborative Intelligence via Cloud-Native Satellites}
\author{Shangguang Wang\inst{1,2}, Qiyang Zhang\inst{1}, Ruolin Xing\inst{1}, Fei Qi\inst{3}, Mengwei Xu\inst{1,*}\corinfo{mwx@bupt.edu.cn}}
\receiveddate{Jun. 25, 2022}
\reviseddate{XX. XX, 2022}
\Editor{XX}

\address[1]{State Key Laboratory of Networking and Switching, Beijing University of Posts and Telecommunications, Beijing 100876, China}
\address[2]{Star Network and Intelligent Computing Laboratory, Beiyou Shenzhen Institute, Shenzhen 518000, China}
\address[3]{Huawei Technologies Co., Ltd}

\begin{document}

\maketitle

\begin{abstract}
Recent advancements in satellite technologies and the declining cost of access to space have led to the emergence of large satellite constellations in Low Earth Orbit. However, these constellations often rely on bent-pipe architecture, resulting in high communication costs. Existing onboard inference architectures suffer from limitations in terms of low accuracy and inflexibility in the deployment and management of in-orbit applications. To address these challenges, we propose a cloud-native-based satellite design specifically tailored for Earth Observation tasks, enabling diverse computing paradigms. In this work, we present a case study of a satellite-ground collaborative inference system deployed in the Tiansuan constellation, demonstrating a remarkable 50\% accuracy improvement and a substantial 90\% data reduction. Our work sheds light on in-orbit energy, where in-orbit computing accounts for 17\% of the total onboard energy consumption. Our approach represents a significant advancement of cloud-native satellite, aiming to enhance the accuracy of in-orbit computing while simultaneously reducing communication cost.
\end{abstract}

\section{Introduction}
\label{Introduction}
With the evolution of next-generation communication networks and advanced technologies, 
the number of end-user connections and application requirements are increasing.
It is estimated that two-thirds of the population in the world will have access to the Internet and billions of devices will be connected to the Internet by 2023 \cite{Cis}.
However, there are still more than 80\% of the land and 90\% of the ocean on the Earth without network coverage.
Moreover, real-time and high-resolution satellite imagery have vital
use in getting along with nature. The terrestrial network is extremely vulnerable in the face of natural disasters, such as earthquakes and floods.

In recent years, however, the increasing investments in satellite-related technologies make the Low Earth Orbit (LEO) satellite constellation a strong complement to terrestrial networks.
Numerous competitors have disclosed efforts to deploy LEO constellations, including SpaceX \cite{space}, Telesat \cite{te}, and Amazon \cite{am}. 
These LEO constellations are promising to blanket the globe with low-latency broadband Internet due to the declining costs of satellite manufacturing and launching today.
Satellites are typically categorized into distinct types, each specifically designed to fulfill various tasks such as communication, guidance, and sensing. In this work, we focus on the Earth Observation satellites.

\begin{figure}
    \centering
    \includegraphics[width=0.45\textwidth]{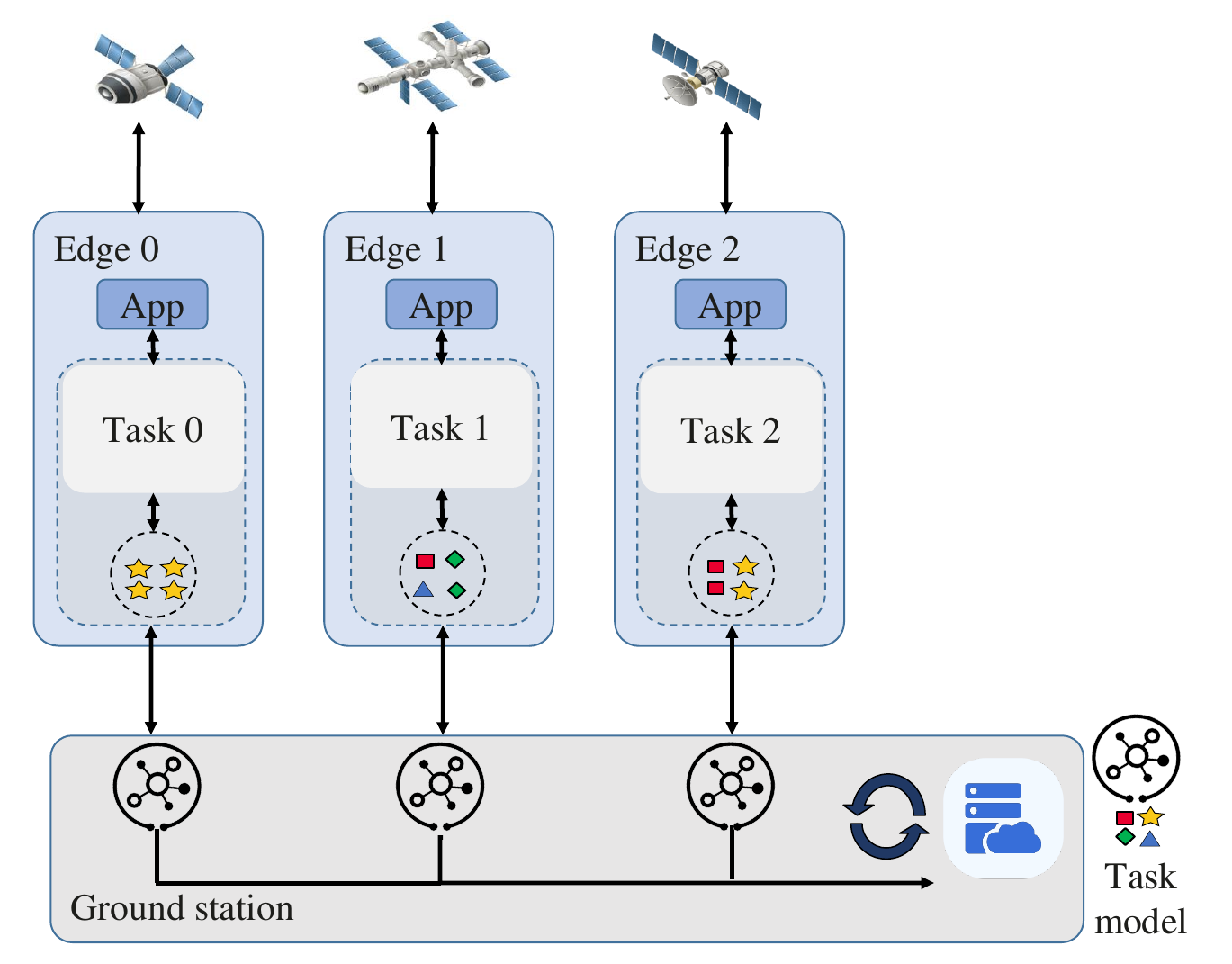}
    \caption{The framework of collaborative artificial intelligence. Cloud-native satellite can support diverse tasks.}
    \label{fig1}
\end{figure}

Since bent-pipe satellites do not have in-orbit processing capabilities, raw data is completely transmitted to the ground station, incurring a communication cost \cite{denby2020orbital}.
The architecture breaks down due to the limitations on downlink availability.
Compression is useful because it reduces the resources required to store and transmit data. However, computational resources are consumed in compression and decompression.
Orbital Edge Computing (OEC) is proposed to reduce the downlink transmission by placing computing resources at the LEO satellites constellation \cite{2019Satellite}. 
A promising cloud-native satellite operates by capturing an image and processing it locally instead of transmitting it to the ground completely.
Though the satellite with computing capability has been developed \cite{zhong2022machine}, satellite-ground collaborative intelligence is still required as follows:   
it is difficult for satellites to perform complex tasks independently, despite their limited computing capability;
the models are one-size-fits-all for all scenarios because satellite data are inconsistently in spatial and temporal distribution.
Based on these reasons, satellite-ground collaborative intelligence still plays a vital role now.

Cloud-native satellite technologies show great promise for in-orbit AI applications
when the real-world deployment brings forth a set of unique challenges.
One such challenge is the lack of flexibility in existing onboard inference architectures, which hinders the efficient deployment and management of in-orbit applications. Additionally, these architectures suffer from low accuracy levels that fail to meet the demanding requirements of such applications.
To address these challenges, we propose a cloud-native-based collaborative intelligence architecture designed to cater to various in-orbit AI applications, including CNN-based image classification, object detection, and segmentation.
By adopting this architecture, we aim to enhance the adaptability and accuracy of in-orbit AI applications, paving the way for their successful implementation in a real-world context.

We introduce a case study of satellite-ground collaborative inference system deployed in the experimental satellites of Tiansuan constellation, Baoyun and Chuangxingleishen. Our system has yielded a remarkable 50\% improvement in in-orbit inference accuracy and a substantial 90\% reduction in the amount of data returned by the satellite. However, in-orbit processing must be compatible with the space environment, which poses challenges due to the higher cost and lower capabilities of space-based systems compared to corresponding COTS units on the ground.
Furthermore, we report our findings regarding the energy consumption of in-orbit computing, which accounts for approximately 17\% of overall energy consumption. These observations have important implications for future satellite designs and energy management strategies. Our work represents a significant step towards realizing the potential of cloud-native satellites for improving the efficiency and accuracy of in-orbit computing.

The main contributions of this paper as follows:
\begin{itemize}
  \item [1)] We propose the concept of cloud-native satellites to cater to in-orbit computing and facilitate promising applications.       
  \item [2)] We present a case study of a satellite-ground collaborative inference system deployed in Tiansuan constellation. 
  \item [3)] We report that the inference system not only enhances the accuracy of in-orbit computing but also effectively reduces communication costs. Additionally, we provide insights into the energy consumption associated with in-orbit computing.
\end{itemize}

The rest of this paper is organized as follows. Section \uppercase\expandafter{\romannumeral2} introduces the background and motivation of the paper.
Section \uppercase\expandafter{\romannumeral3} describes the overall system architecture and main components.
Then Section \uppercase\expandafter{\romannumeral4} details how the satellite-cloud collaborative inference works.
Finally, Section \uppercase\expandafter{\romannumeral5} concludes the paper and discusses the potential directions.

\section{Background \& Motivation}

\textbf{Satellite data is exploding.}
It's reported that about 45\% of the LEO satellites in orbit are used for earth observation.
The data amount obtained by onboard cameras is too large and 60\% of remote sensing images are highly similar in the same scenario \cite{avtar2022multi}.
For example, ZY-3 is the Chinese first civilian high-precision cartographic satellite, and it generates more than 10 of TB data every day \cite{ZY}.
One single remote sensing image dataset can reach tens of GB \cite{ground}.
Such large data transmission to the ground station may take up most of the time and induce huge communication overhead under the bent-pipe architecture \cite{park2019wireless}.
It is a fact that not all raw observations are worth exploring further.
Especially in the southwest of China, 80\%-90\% of raw data is invalid due to cloud cover. 
There has been preliminary research in the field of onboard image preprocessing for remote sensing satellites. 
Before satellites backhaul, redundant information such as cloud cover area can be eliminated in advance and the data returned can be greatly reduced.

\textbf{Downlinks can be unreliable.}
Existing systems under a bent-pipe architecture stem from fundamental physical constraints simply downlink raw observations to ground stations.
Surprisingly, one satellite task lost 80\% of its data packets due to downlink instability \cite{nogales2018makersat}.
The time-varying relationship between the orbital position of the satellite and the geographic location of ground stations imposes limitations on link availability
and can incur high downlink latency between data collection and processing. 
Any viable orbital system must directly address the unique physical constraints.
As an alternative paradigm, OEC colocates processing hardware with high-datarate sensors in low-cost satellites.

\textbf{Onboard processing is on the rise.}
With the development of onboard chip integration, the processing performance of satellites has been improved.
Onboard systems have integrated multiple computing modules such as CPUs, GPUs, DSPs and FPGAs \cite{schwenk2018scosa}.
Different from traditional CPUs, DSPs and GPUs have stronger computing capabilities.
For example, based on embedded cluster computing, the next-generation onboard computer makes a system architecture 
that combines the master nodes (e.g., CPUs) and slave nodes (e.g., GPUs, FPGAs and DSPs) for parallel processing \cite{ludtke2014obc}.
These processing units can work together to provide higher onboard computing and improve reconfiguration capabilities.
In a word, the enhancement of onboard computing processing capability lays the foundation for OEC.

\textbf{Enhancing OEC Landing with Cloud-Native Technology.}
While computing resources at the LEO satellite constellation enables OEC, meeting the demand for flexible deployment in orbit remains a significant challenge.
Cloud-native techniques with container and microservice empowers satellites to build and run scalable applications in a modern, dynamic environment.
These techniques facilitate the development of loosely coupled systems that exhibit resilience, manageability, and observability.
Cloud-native satellite overcomes challenges for unprecedented success in real-world deployment.

Previous works related to this topic in the network/system community have been proposed. Denby et al. \cite{denby2020orbital} propose an orbital edge computing architecture on satellite constellations to reduce system edge processing latency. Vasisht et al. \cite{vasisht2021l2d2} present a scheduler for multi-satellite, multi-ground station configuration to reduce data downlink latency. Jinhyun et al. \cite{so2022fedspace} focus on the federated learning framework, which dynamically schedules model aggregation based on the deterministic and time-varying connectivity according to satellite orbits. \cite{cui2022space, xiao2022leo} also present challenges of space-air-ground integrated network architecture.
\cite{kato2019optimizing, furano2020towards} discuss opportunities and potential application areas for intelligence in space systems.
Our work is orthogonal to and compatible with those above works.
To our knowledge, we present a novel approach to achieve high inference accuracy through satellite-ground collaborative intelligence using cloud-native satellites. 
Our work marks the effort in harnessing the full potential of cloud-native satellite, thereby paving the way for enhanced efficiency and accuracy in in-orbit computing.

\section{Design of Cloud-native Satellites}
With the expansion of LEO constellations and the explosive growth of satellite data, it is urgent to explore how to effectively utilize local satellite data.
Additionally, local processing of satellite data is promising as AI capabilities migrate from the cloud to edge nodes \cite{zhang2022cost}.
However, each satellite has limited computing capability and energy resources, making it difficult to perform complex tasks independently.
Satellite-ground collaboration is proposed to facilitate intelligent in-orbit computing. Also,
motivated by cloud-native,
we first provide an overview of an individual cloud-native satellite.
We then describe key techniques involved in cloud-native satellite, which organizes
a system that supports in-orbit computing.

\subsection{Cloud-native satellite}
As the first cloud-native satellite in the world, the satellite leverages advanced techniques such as containers and microservices to realize in-orbit computing.
As shown in Figure 2, we integrate communication, computing, and storage capabilities into cloud-native satellites. The satellite provides open interfaces 
that allow third parties to deploy applications. The satellite also monitors and manages the operational status and applications.
Based on the storage capacity, computing module provides onboard control capabilities for basic and third-party ubiquitous applications.
Cloud-native satellites guarantee in-orbit service execution that mainly benefits from:

\begin{figure}
    \centering
    \includegraphics[width=0.45\textwidth]{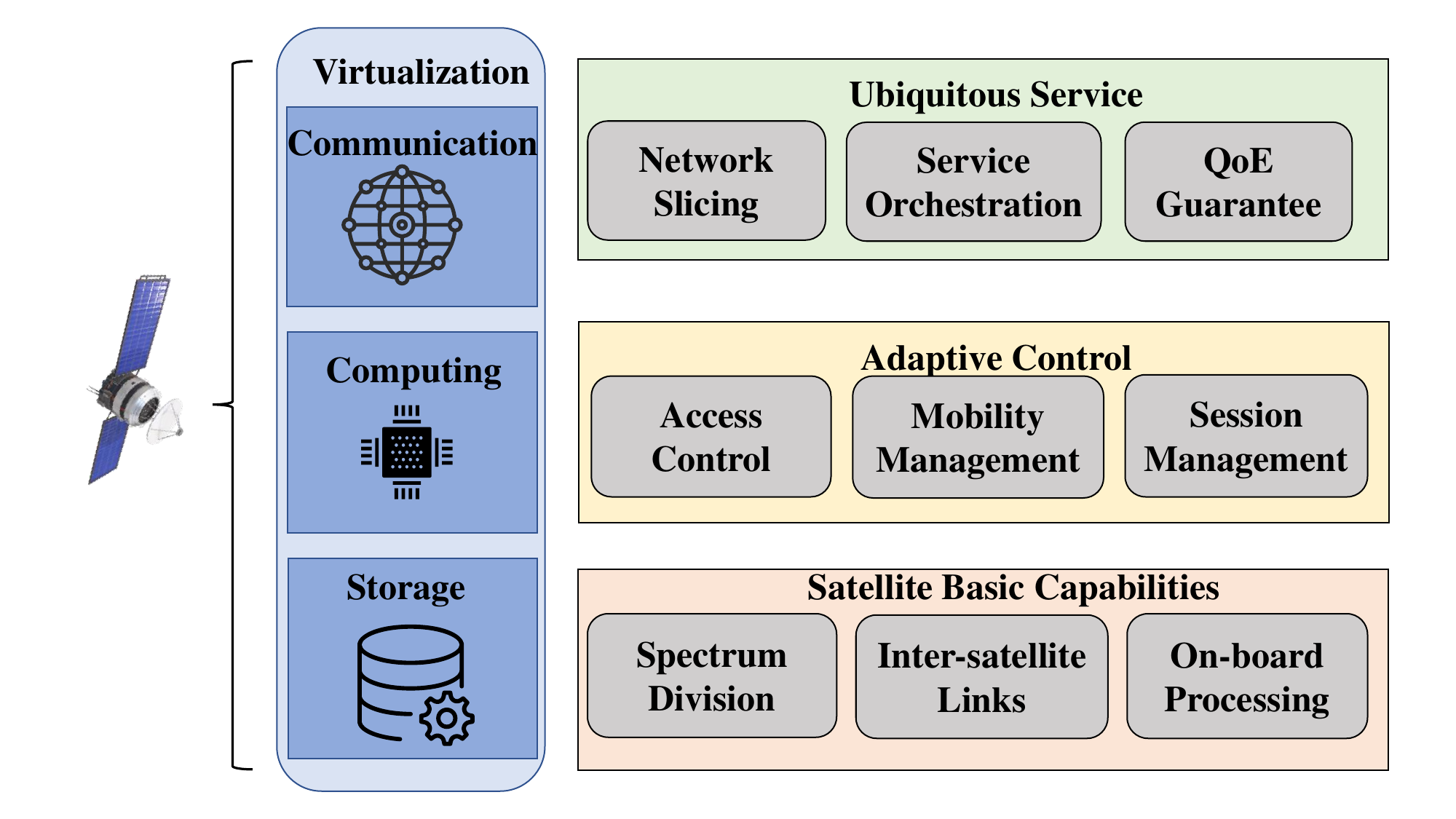}
    \caption{The design of the cloud-native satellite. The communication, computing, and storage capabilities are integrated into the cloud-native satellite.}
	\label{fig1}
\end{figure}

\begin{figure*}
    \centering
    \includegraphics[width=\textwidth]{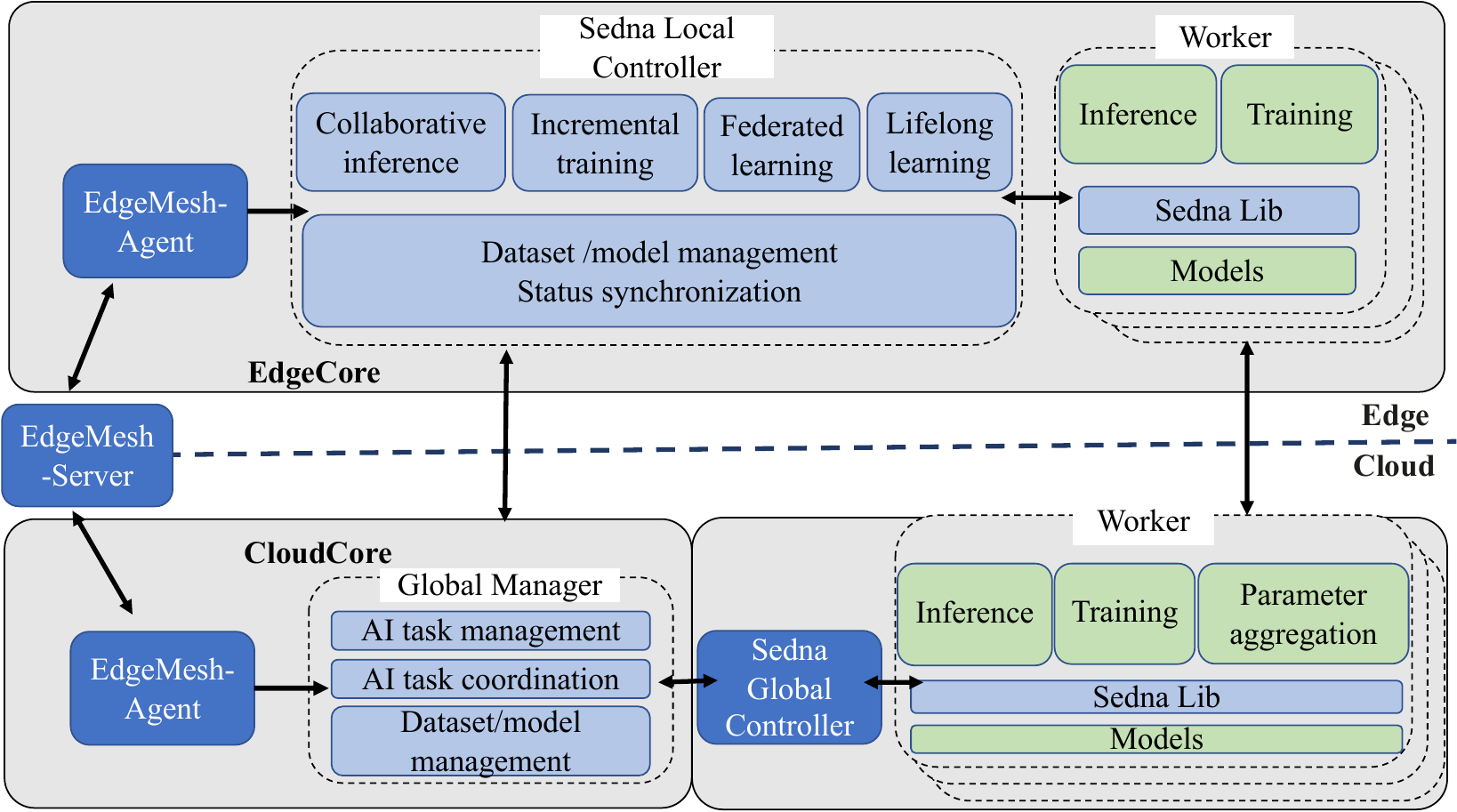}
    \caption{The system architecture of satellite-ground collaborative based on KubeEdge.}
	\label{fig1}
\end{figure*}

\textbf{Rapid deployment}
Cloud-native satellite leverages agile development and DevOps project management models \cite{damiani2022blastfunction}.
It not only enables users to deploy applications quickly and automatically but also continuously updates onboard applications to keep up with changes in demand.

\textbf{Elastic expansion}
The satellite manages satellite-ground clusters by automatically increasing or decreasing edge nodes.
It also shrinks or expands resources according to the real-world situation to improve the utilization of onboard resources.

\textbf{Highly autonomous}
Based on container technology, cloud-native satellite encapsulates service functions to achieve isolation and abstraction independent of location and environment.
Moreover, EdgeMesh, as the data plane component of the KubeEdge cluster, provides simple service discovery and traffic proxy functions for satellite service, 
thereby shielding the complex network structure in edge scenarios \cite{edge}.

\textbf{Safe and reliable}
Container orchestration is crucial for large and dynamic production environments \cite{zhong2022machine}.
With container orchestration for microservices, onboard applications can be automatically scaled, fault-tolerant 
which copes with the complex environment of space and keeps onboard applications available at all times.

The emergence of cloud-native satellites benefits from the application management framework in which satellites and the ground cloud work together.
KubeEdge \cite{kubeedge} and its sub-project Sedna \cite{Sedna} provide unified resource and application management for cloud-native satellites and improve application collaboration capabilities.
The system architecture is shown in Figure 3.
Collaborative AI framework breaks model training and inference space constraints.
In the following, we mainly introduce the two major frameworks: KubeEdge and Sedna.

\subsection{KubeEdge}

KubeEdge is an open-source platform based on Kubernetes that provides infrastructure support for applications by extending containerized application orchestration to satellites \cite{kubeedge}.
KubeEdge also synchronizes deployments and metadata between central cloud and satellite, which strongly supports:

\textbf{Extremely lightweight}
The global resources of the satellites are discrete and the local resources are limited to one satellite.
The instability of the edge network leads to frequent disconnection of edge satellites and management systems.
The platform extracts the core functions of kubelet and builds a lightweight management software named EdgeCore to cover different scenarios.
EdgeCore and CloudCore in the central cloud provide data centers to meet exacting requirements without compromise \cite{edgecore}.

\textbf{Reliable connection}
The network between satellites and ground station often suffers from low bandwidth and serious packet loss.
The platform manages edge-cloud messages in the same way, and the data is still reliably transmitted in weak network scenarios.

\textbf{Offline autonomous}
A lightweight management component named MetaManager stores metadata \cite{rani2022interface}. 
When edge nodes go offline, applications are managed and restored based on storage metadata.
This method not only brings less resource overhead but also ensures the security of other satellite nodes in the cluster.

\textbf{Cloud-edge collaboration}
Edge and cloud always work together to achieve end-to-end applications.
KubeEdge provides an edge-cloud data exchange mode. 
More importantly, EdgeMesh provides unified service discovery and traffic proxying between microservices.
For example, the capabilities of EdgeMesh-Server are merged into the tunnel module of EdgeMesh-Agent, so that EdgeMesh-Agent with relay capability can automatically become a relay server, providing other nodes with the functions of assisting hole punching and relaying.

\subsection{Sedna}
Sedna's edge-cloud synergy is implemented for collaborative inference based on the capabilities provided by KubeEdge. The component Global Manager in Figure 3 unifies satellite-ground synergy AI task management and collaboration by well-supporting current popular frameworks including TensorFlow/ Pytorch/ PaddlePaddle/ MindSpore, etc \cite{Sedna}.
Sedna helps AI applications migrate seamlessly to satellites by using cloud-edge collaboration protocols to reduce cost, improve model performance, and protect data privacy, etc.
In the following section, we discuss the major components of Sedna.

\textbf{Global Manager}
This component implements an edge AI controller based on Kubernetes.
Users create CRD \cite{crd} to achieve model/dataset management, AI task management for edge-cloud collaboration, and edge-cloud coordination.

\textbf{Local Controller}
This component performs local process control of edge-cloud collaborative AI tasks.
Moreover, it also performs local general management: models, datasets, state synchronization of AI tasks, etc.

\textbf{Worker} 
This component performs AI tasks based on the training/ inference procedures of existing AI frameworks. 
Workers can be deployed on the edge or in the cloud and they work together with each other.

\textbf{Lib} This component provides an edge-cloud collaboration interface for applications. Users can implement training/ inference and aggregation based on this Lib.

\begin{figure}
    \centering
    \includegraphics[width=0.45\textwidth]{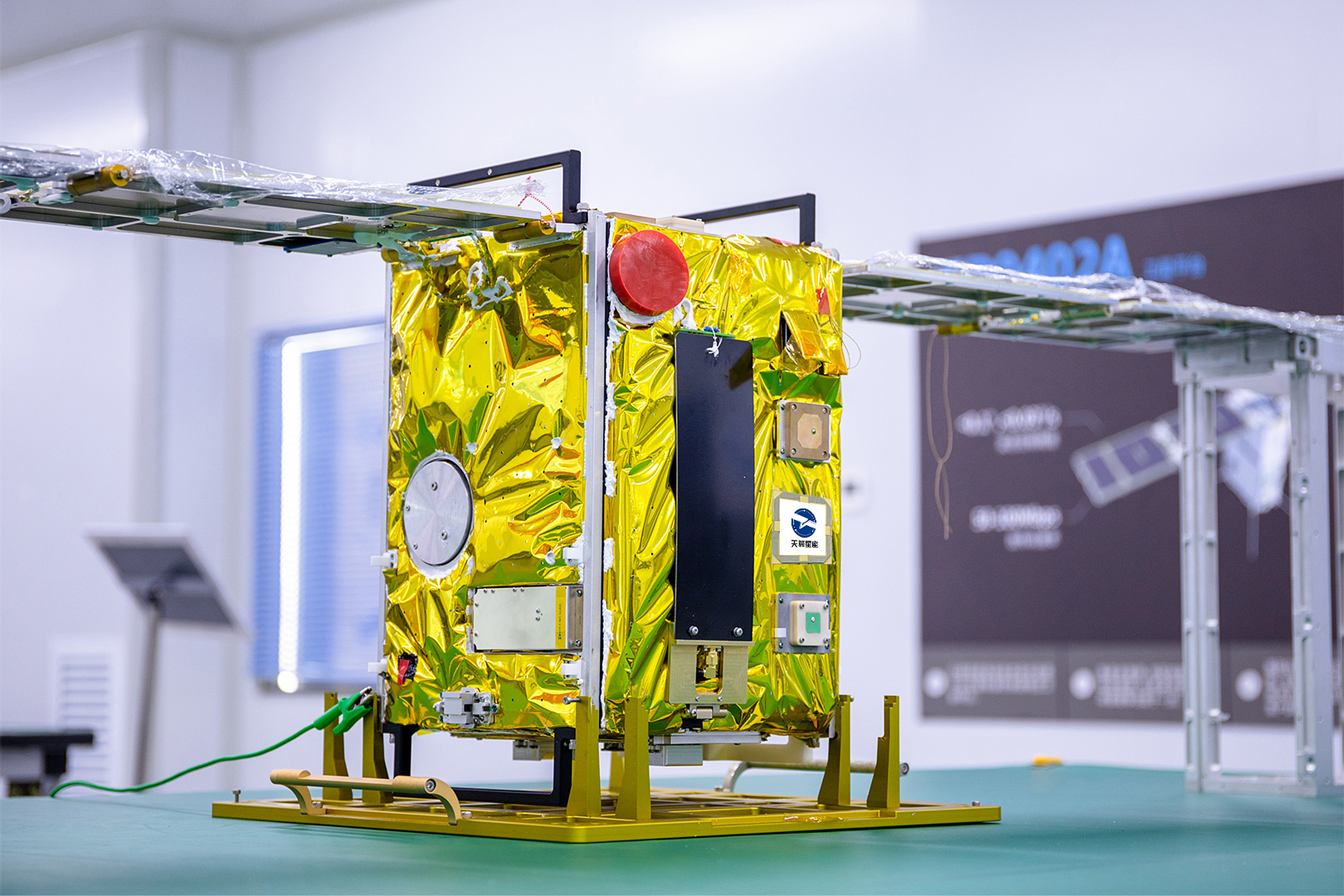}
    \caption{A physical appearance of the Baoyun satellite.}
	\label{fig1}
\end{figure}

\subsection{Promising Application scenarios}
To address different tasks and challenges, cloud-native satellites can support of diverse paradigms. Currently supported computing protocols are as follows:


\textbf{Collaborative inference} 
Satellites cannot independently perform complex inference tasks because of limited resources and weak computing capability on the satellite. 
The satellite performs one part of the whole inference and offloads some tasks to the cloud to improve the overall inference performance.

\textbf{Incremental training} 
Changes in data distribution, such as weather, make the deployed model unable to adapt to the current environment and the detection accuracy decreases.
For this challenge, satellites continuously collect newly generated data and train models in the cloud.
The satellite nodes regularly fine-tune the model from the cloud to improve accuracy.

\textbf{Federated learning} 
From the perspective of privacy protection, federated learning can effectively alleviate data silos because raw data is reluctant to share to the cloud.
The satellite trains the model and transmits the parameters (i.e., training weights) to the cloud responsible for parameter aggregation.
The system also provides a secure data channel through KubeEdge and supports encrypted transmission to ensure data security.
Such a protocol may train high-precision models for privacy protection.

\begin{figure*}
    \centering
    \includegraphics[width=\textwidth]{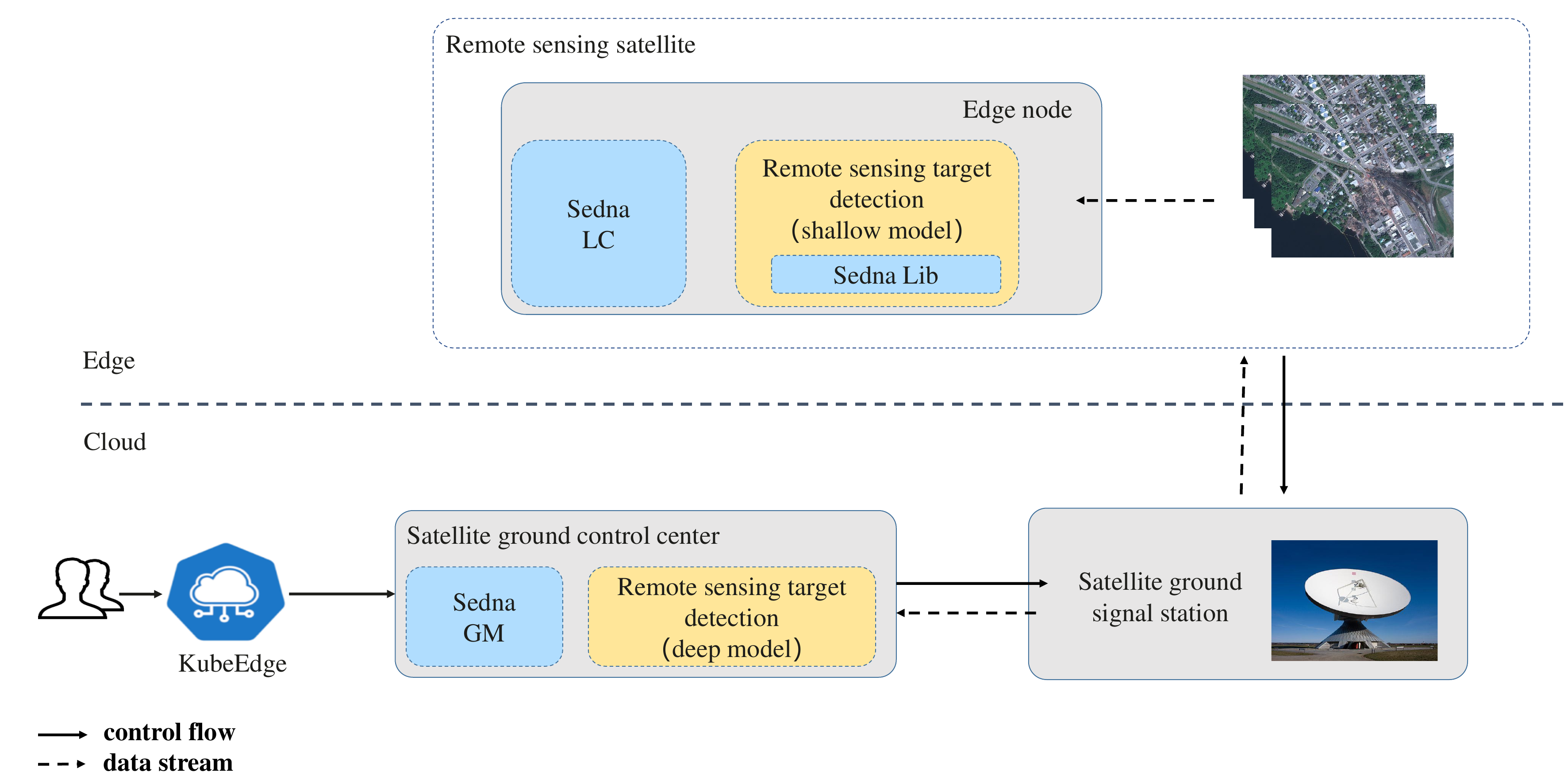}
    \caption{The workflow of satellite-ground collaborative inference system architecture.}
	\label{fig1}
\end{figure*}

\textbf{Lifelong learning} 
Satellites suffer from data drift and catastrophic forgetting of onboard models.
Combining incremental training and multi-task training, satellite model enables knowledge transfer across time and scenarios.
Based on the knowledge library in the cloud, satellite model can be continuously updated to address unknown tasks.

\section{Case Study: The First in-orbit AI Collaborative Inference}

We've successfully launched two new satellites, Baoyun and Chuangxingleishen, and implemented collaborative inference on these satellites using KuebeEdge and Sedna. Earth Observation satellites incorporate cloud-native computing capabilities.
Table 1 shows the real-world specifications including vital parameters such as mass, size for both satellites. Additionally,
Figure 4 provides a physical appearance of Baoyun satellite. It's worth noting that these satellites adhere to the CubeSat standard for design.
The control center operates on a Linux server, sending remote instructions to the satellites. We've deployed two different precision detection models, consisting of lightweight, low-precision models onboard the satellite for preprocessing to obtain preliminary results. On the ground, large, high-precision models are situated with high computing capabilities, which facilitate accurate detection.
The workflow of our collaborative inference system is shown in Figure 5. The satellites perform a preliminary detection using a lightweight model to identify features of interest. If confidence threshold in the results is high, the processed results are sent back to the ground directly. However, if confidence threshold is low, the satellite transmits the images to the ground, where the high-precision detection model is used for exact detection. This system enables efficient detection of captivating images and information, ensuring that only essential data is transmitted to the ground.
Our experiments are based on the fact that onboard inference can be performed anytime, when satellite computation are available. The handover between them only occurs during the contact time between the satellite and the ground.
Note that the shown evaluations are conducted in Baoyun satellite. Interestingly, the experimental results on different satellites show that the inference system can achieve consistent results on satellites of different specifications.

Onboard image splitting significantly reduces the occurrence of redundant images caused by cloud cover.
We conducted experiments on the widely-used DOTA \cite{dota}, object detection dataset. 
We propose a strategy to split the images into smaller images before performing in-orbit inference.
This approach is necessary due to the limited computing power of the satellite, which cannot handle high-resolution images.
Figure 6 illustrates the reduction in redundant images during in-orbit operations. We observe that splitting the large images into smaller fragments, irrespective of the fragment size, leads to a remarkable 90\% and 40\% decrease in images for the two versions of the datasets, respectively. As a result, in-orbit processing approach 
offers the dual benefits of reducing the number of images and conserving significant bandwidth.
\renewcommand\arraystretch{2}
\begin{table*}[]
\centering
    \caption{Satellite platform specifications.}
    \begin{tabular}{lllllllll}
    \hline
    Name & \makecell[c]{Launch \\Time}    & \makecell[c]{Orbital \\Altitude\\(km)} & \makecell[c]{Mass\\(kg)} & \makecell[c]{Load \\Size\\(U)} & \makecell[c]{Size \\(U)}& \makecell[c]{Operating \\System} & \makecell[c]{Uplink \\Rate\\(Mbps)} & \makecell[c]{Downlink \\Rate\\(Mbps)} \\ \hline
    Baoyun            & \makecell[c]{Dec. 7 \\2021}  & 500±50    & 20 & \makecell[c]{0.25} & \makecell[c]{12} &\makecell[c]{Ubuntu Server \\20.04 arm} & 0.1$\sim$1 & $\geq$40       \\
    \makecell[c]{Chuangxing\\leishen} & \makecell[c]{Feb. 27 \\2022} & 500±50    & 20 & \makecell[c]{0.25} & \makecell[c]{6} &\makecell[c]{Debian Buster with \\Raspberry Pi} & 0.1$\sim$1 &  $\geq$40     \\ \hline
    \end{tabular}    
\end{table*}
Collaborative inference yields a significant improvement in accuracy.
Our satellite-ground collaborative inference system incorporates YOLOv3-tiny and YOLOv3 object detection models for onboard and ground detection, respectively. To assess the accuracy of our system, we use the mean average precision (mAP) metric, which compares ground-truth bounding boxes to detected boxes and returns a score. A higher score indicates more accurate detection \cite{ren2015faster}.
We analyzed the accuracy of our system through in-orbit and collaborative inference methods, as shown in Figure 7.
The results demonstrate a remarkable accuracy enhancement of 44\% and 52\% respectively when employing collaborative inference compared to in-orbit inference. On average, our collaborative inference system achieves an approximate 50\% improvement in accuracy.
Additionally,
a significant number of redundant images were detected. By performing preprocessing onboard and directly transmitting the inference results to the ground station, our system successfully reduced the amount of data returned by 90\%.
\begin{figure}
    \centering
    \includegraphics[width=0.45\textwidth]{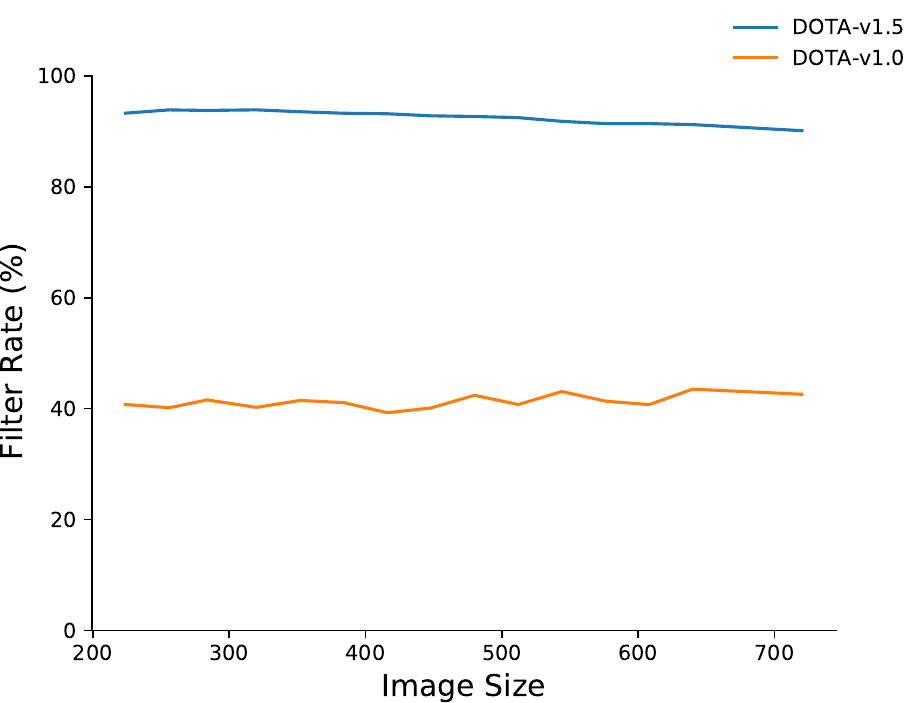}
    \caption{The filter rate of redundant data in orbit on DOTA.}
	\label{fig1}
\end{figure}

\begin{figure}
    \centering
    \includegraphics[width=0.45\textwidth]{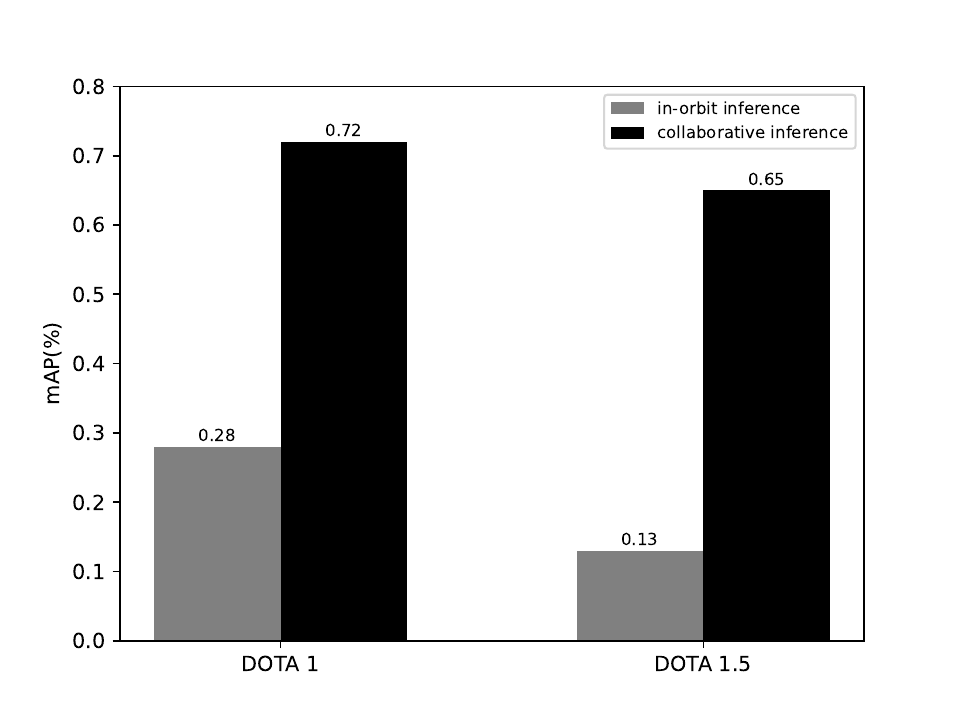}
    \caption{Accuracy (mAP in object detection task) of in-orbit vs. collaborative inference.}
	\label{fig1}
\end{figure}

According to the energy consumption analysis presented in Table 2, Baoyun satellite's energy system primarily powers its payloads, accounting for approximately 53\% of the total energy consumption. The remaining energy is utilized for basic satellite functions such as communication and propulsion.
Further delving into the energy consumption of the satellite's payloads, 
Onboard equipment measures the voltage and current of each power system and records the telemetry data, which is then transmitted to the ground. The reported power consumption is recalculated based on the collected data. Table 3 presents a breakdown of the subsystems responsible for satellite computing and their corresponding power consumption, including the camera that captures remote sensing images. 
Among these power subsystems, the Raspberry Pi system accounts for the largest proportion of energy consumption, representing 33\% of the total energy consumed by the payloads.
In summary, computing on the Baoyun satellite accounts for about 17\% of energy consumption.
The presented data sheds light on the energy allocation and consumption patterns of the satellite, offering value for optimizing its operational efficiency and ensuring its long-term sustainability.

\renewcommand\arraystretch{2}
\begin{table*}[]
    \centering
    \caption{Real power distribution of energy consumption system in Baoyun satellite.}
    \begin{tabular}{llllllll}
    \hline
    \makecell[c]{Item}             & \makecell[c]{Electrical \\Subsystem}    & Propulsion  & Guidance &Avionics & Comm.  & Payloads& Sum  \\ \hline
   \makecell[c]{Power(W)}           & \makecell[c]{1.47}  & \makecell[c]{7.00}       & \makecell[c]{5.43}     & \makecell[c]{4.81}     & \makecell[c]{5.43}  & \makecell[c]{26.93} & \makecell[c]{51.07}   \\ \hline     
    \end{tabular}
    
\end{table*}

\begin{table*}[]
    \centering
    \caption{The power of payloads subsystem of Baoyun satellite.}
    \begin{tabular}{llllllll}
    \hline
    \makecell[c]{Item}             & \makecell[c]{Camera}    & Occultation  & Tribology  & Mems & Adsbs& Raspberry Pi \\ \hline
   \makecell[c]{Power(W)}           & \makecell[c]{0.09}  & \makecell[c]{6.26}       & \makecell[c]{5.68}     & \makecell[c]{0.95}     & \makecell[c]{6.12}  & \makecell[c]{8.78}   \\ \hline     
    \end{tabular}
\end{table*}

\section{conclusion\& Future Work}
This paper introduces the emergence of a cloud-native satellite that directly leverages onboard computering for remote sensing tasks.
It explores various execution protocols such as collaborative inference, federated learning, incremental learning, and lifelong learning.
Additionally, we launched the first cloud-native satellite and implemented the collaborative inference system on the satellite,
with the ultimate objective of achieving high inference accuracy through the use of cloud-native satellites.

In the future, we plan to deploy this project in batches across other satellites of Tiansuan constellation, thereby establishing a collaborative computing network in space.    
We also expect cloud-native satellites to unlock new capabilities in precision agriculture, infrastructure monitoring, humanitarian assistance, etc \cite{lang2020earth}.

\section{ACKNOWLEDGEMENT}
This work was supported by National Natural Science Foundation of China (62032003).

\bibliographystyle{gbt7714-numerical}
\bibliography{myref.bib}

\biographies

\begin{CCJNLbiography}{wangshangguang}{Shangguang Wang}
    is a professor at the School of Computer Science, Beijing University of Posts and Telecommunications, China. He is the founder\&chief scientist of Tiansuan Constellation. He is also the director of Star Network and Intelligence Computing Laboratory, and vice director of Sate Key Laboratory of Networking and Switching at BUPT.
    His research interests include service computing, mobile edge computing, cloud computing, and satellite computing. He is currently serving as chair of IEEE Technical Committy on Services Computing(TCSVC), and vice chair of IEEE Technical Committee on Cloud Computing. He also served as general chairs or program chairs of 10+ IEEE conferences, advisor/associate editors of several journals such as Journal of Cloud Computing, Journal of Software: Practice and Experience, 
    International Journal of Web and Grid Services, China Communications, and so on. He is a senior member of the IEEE, and Fellow of the IET.
\end{CCJNLbiography}

\begin{CCJNLbiography}{qiyangzhang}{Qiyang Zhang}
received the bachelor’s degree in Network Engineering from Henan Normal University in 2018. 
Currently, he is a Ph.D. candidate in computer science at the State Key
Laboratory of Networking and Switching Technology,
Beijing University of Posts and Telecommunications.
His research interests include satellite edge computing and edge intelligence.
\end{CCJNLbiography}

\begin{CCJNLbiography}{xingruolin}{Ruolin Xing}
received the bachelor’s degree from Beijing University of Posts and Telecommunications in 2019. 
Currently, he is a Ph.D. candidate in computer science at the State Key
Laboratory of Networking and Switching Technology,
Beijing University of Posts and Telecommunications.
His research interests include satellite edge computing and edge intelligence.
\end{CCJNLbiography}

\begin{CCJNLbiography}{qifei}{Fei Qi}
    is the Principal Engineer of Huawei Cloud Computing Technologies and the team leader of Huawei edge-cloud collaboration OS technology innovation team. He joined Huawei in 2012, he has been engaged in the research of cloud computing systems and key technologies for a long time, 
    and has participated in the innovation and incubation of multiple cloud computing products such as Huawei hybrid cloud and edge cloud.
\end{CCJNLbiography}

\begin{CCJNLbiography}{xumengwei}{Mengwei Xu}
    received the bachelor’s and Ph.D. degrees from Peking University, Beijing, China. He is an Assistant Professor with the Computer Science Department, Beijing University of Posts and Telecommunications, Beijing. 
    His research interests cover the broad areas of mobile computing, edge computing, and operating systems.
\end{CCJNLbiography}

\end{document}